\documentclass{article}
\usepackage{graphicx}
\usepackage{amssymb}
\usepackage{amsmath}

\begin{document}

\newcommand{\del}[1]{\delta^{(2)}\left(#1\right)}
\newcommand{\non}{\nonumber\\}

\newcommand{\bra}[1]{\langle #1|}
\newcommand{\ket}[1]{|#1\rangle}
\newcommand{\braket}[2]{\langle #1|#2\rangle}
\newcommand{\kone}{{\bf k}_1}\newcommand{\ktwo}{{\bf k}_2}
\newcommand{\ddkone}{d^{2}{{\bf k}_1}}\newcommand{\ddktwo}{d^{2}{{\bf k}_2}}
\newcommand{\qone}{{\bf q}_1}\newcommand{\qtwo}{{\bf q}_2}
\newcommand{\ddqone}{d^{2}{{\bf q}_1}}\newcommand{\ddqtwo}{d^{2}{{\bf q}_2}}

\newcommand{\shat}{\hat{s}}
\newcommand{\asbar}{\bar\alpha_s}
\newcommand{\opk}{\hat{\mathcal{K}}}
\newcommand{\opf}{\hat{f}}
\newcommand{\opjetone}{\hat{\Phi}_{\rm jet,1}}
\newcommand{\opjettwo}{\hat{\Phi}_{\rm jet,2}}
\newcommand{\oplept}{\hat{\Phi}_{\rm leptonic}}
\newcommand{\opqone}{\hat{\bf q}_1}\newcommand{\optwo}{\hat{\bf q}_2}

\newcommand{\plusinu}[1]{\left(#1\right)^{i\nu-\frac{1}{2}}}
\newcommand{\minusinu}[1]{\left(#1\right)^{-i\nu-\frac{1}{2}}}
\newcommand{\plusinup}[1]{\left(#1\right)^{i\nu'-\frac{1}{2}}}
\newcommand{\minusinup}[1]{\left(#1\right)^{-i\nu'-\frac{1}{2}}}

\title{\Large Azimuthal decorrelation of forward jets in Deep Inelastic Scattering}
\author{{\sc Agust{\' \i}n~Sabio~Vera}$^1$ and {\sc Florian~Schwennsen}$^2$\\[2.5ex]
$^1$ {\it Physics Department, Theory Division, CERN,}\\{\it CH--1211, Geneva 23, Switzerland}\\
$^2$ {\it II. Institut f\"{u}r Theoretische Physik, Universit\"{a}t Hamburg,}\\
 {\it Luruper Chaussee 149, D--22761 Hamburg, Germany}}

\maketitle

\vspace{-9cm}
\begin{flushright}
{\small CERN--PH--TH/2007--130\\DESY--07--113}
\end{flushright}

\vspace{9cm}
\begin{abstract}
We study the azimuthal angle decorrelation of forward jets in Deep Inelastic 
Scattering. We make predictions for this observable at HERA describing the 
high energy limit of the relevant scattering amplitudes 
with quasi--multi--Regge kinematics together with a 
collinearly improved evolution kernel for multiparton emissions.

\end{abstract}

\section{Introduction}

In previous publications~\cite{Vera:2006un,Vera:2007kn,Schwennsen:2007hs} 
we have studied the 
azimuthal angle decorrelation between Mueller--Navelet jets at different 
hadron colliders within the Balitsky--Fadin--Kuraev--Lipatov (BFKL) 
formalism~\cite{BFKL} beyond leading order accuracy~\cite{NLA-kernel}. 
In the present work we extend these studies to predict the decorrelation in 
azimuthal angle between the electron and a forward jet associated to the 
proton in Deep Inelastic Scattering (DIS). When the separation in rapidity 
space, $Y$, between the scattered electron and the forward jet is large and 
the transverse momentum of the jet is similar to the virtuality of the photon 
resolving the hadron, then the dominant terms in the scattering amplitude are 
of the form $\sim (\alpha_s Y)^n$. These terms can be resummed to all orders 
by means of the BFKL integral equation. The calculation for this process 
is very similar to that of Mueller--Navelet jets, the only difference being 
the substitution of one jet vertex by a vertex describing the 
coupling of the electron to the NLO BFKL gluon Green's function via a 
quark--antiquark pair.

This observable was previously investigated in the leading logarithmic 
approximation (LO) in Ref.~\cite{Bartels:1996wx}. In the following we build 
on that work while still using the LO approximation for the jet vertex and the 
virtual photon impact factor. However, we improve the calculation by 
considering 
the BFKL kernel to next--to--leading (NLO) accuracy, {\it i.e.} we also keep 
$\sim \alpha_s (\alpha_s Y)^n$ terms in the gluon Green's function, a 
process--independent quantity which governs the dependence on $Y$ of the 
cross section, previous analysis in this direction can be found in 
Ref.~\cite{Kepka}. The NLO terms in the impact factors should be relevant only in 
the regions of moderate $Y$, to include them in our analysis one could use 
Ref.~\cite{ImpactFactorsAll} for the leptonic part and 
Ref.~\cite{NLOjetvertex} for the forward jet vertex. We will leave this task 
for future work.

\section{Forward jet cross section}

For the production of a forward  jet in DIS it is necessary to extract 
a parton with a large longitudinal momentum fraction $x_{\rm FJ}$ from the 
proton. When the jet is characterized by a hard scale it is possible to use 
conventional collinear factorization to describe the process. Consequently, 
the jet production rate may be written as
\begin{equation}
  \sigma(s) = \int dx_{\rm FJ}\;f_{\rm eff}(x_{\rm FJ},\mu_F^2)\hat\sigma(\shat),
\end{equation}
with $\hat\sigma(\shat)$ denoting the partonic cross section, and the 
effective parton density~\cite{Combridge:1983jn} being 
\begin{equation}
  \label{eq:feff}
  f_{\rm eff}(x,\mu_F^2) = G(x,\mu_F^2)+\frac{4}{9}\sum_f\left[Q_f(x,\mu_F^2)+\bar{Q}_f(x,\mu_F^2)\right],
\end{equation}
where the sum runs over all quark flavors, and $\mu_F$ stands for the 
factorization scale. 

\begin{figure}[htbp]
  \centering
  \includegraphics[width=6cm]{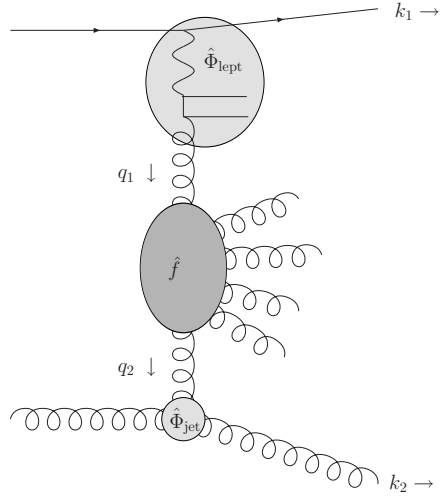}
  \caption{Kinematics of the partonic cross section.}
  \label{fig:herakinematics}
\end{figure}

At partonic level we show a typical configuration contributing to 
$\hat\sigma(\shat)$ in Fig.~\ref{fig:herakinematics}. At the leptonic vertex, 
we treat the quark--antiquark pair inclusively, while we focus on the outgoing 
electron, which carries momentum $k_1$, and the gluon, which couples to the 
Green's function with momentum $q_1$. In our notation the azimuthal angle of 
$k_1$ is $\alpha_1$ and that of $q_1$ is $\theta_1$. 

We also work with commonly used DIS variables such as the proton momentum $P$,
the photon's momentum $q_\gamma$, its virtuality $Q^2=-q_\gamma^2$, the 
Bjorken scaling variable $x_{\rm Bj}=\frac{Q^2}{2Pq_\gamma}$ and the 
inelasticity $y=\frac{Pq_\gamma}{P(q_\gamma+k_1)}$. Making use of the relation 
$\kone^2=(1-y)Q^2$ and the specific structure of the leptonic vertex, 
we can write the partonic cross section in the form
\begin{eqnarray}
  \label{eq:oppartonicep}
  \hat\sigma(\shat) &=& \frac{\pi^2\asbar^2}{2}\int\ddkone\int\ddktwo
\int\frac{d\omega}{2\pi i} e^{\omega Y} 
\bra{\kone} \oplept \opf_\omega \hat{\Phi}_{\rm jet} \ket{\ktwo}
\end{eqnarray}
where the rapidity is defined as $Y=\ln x_{\rm FJ}/x_{\rm Bj}$. In bold we denote the 
transverse Euclidean momenta. We can further decompose the integration and write
\begin{eqnarray}
\hat\sigma(\shat) &=& \frac{\pi^2\asbar^2}{2}\sum_{n,n'=-\infty}^\infty
\int d\alpha_1\int dy \int\ddktwo\int\frac{d\omega}{2\pi i}\int\ddqone\int\ddqtwo 
\int  d\nu\int d\nu' \nonumber\\
&& \hspace{-1cm}\times \bra{y,\alpha_1} \oplept\ket{\qone}\braket{\qone}{\nu,n}\bra{n,\nu}\opf_\omega\ket{\nu',n'}\braket{n',\nu'}{\qtwo}\bra{\qtwo}\hat{\Phi}_{\rm jet}\ket{\ktwo} e^{\omega Y}.
\label{eq:oppartonicep2}
\end{eqnarray}
We have introduced a Fourier expansion on conformal spins $n$, to be defined below. 
The integrals in transverse momenta are taken over the whole two dimensional space while 
the $\nu$ integrations go from $-\infty$ to $\infty$. The contour in the 
$\omega$--plane is to be taken to the right of all possible singularities.

In Eq.~\eqref{eq:oppartonicep2} we have used the transverse momentum representation 
defined by
\begin{align}
  \opqone\ket{\qone} =& \qone\ket{\qone} & \braket{\qone}{\qtwo} =& \del{\qone-\qtwo},
\end{align}
where the kernel in operator form $\opk$, 
\begin{equation}
  \mathcal{K}(\qone,\qtwo) = \bra{\qone}\opk\ket{\qtwo},
\end{equation}
defines the BFKL integral equation at NLO, {\it i.e.},
\begin{align}
  \label{eq:opbfkl}
  \omega\,\opf_\omega =& \hat{1}+\opk\opf_\omega.
\end{align}
To change representation we introduce the basis
\begin{equation}
  \braket{\qone}{\nu,n}=\frac{1}{\pi\sqrt{2}}\plusinu{\qone^2}e^{in\theta_1},
\end{equation}
where $\theta_1$ is the azimuthal angle of $\qone$. The normalization in this new basis reads
\begin{equation}
  \braket{n',\nu'}{\nu,n} = \int\ddqone \frac{1}{2\pi^2}\left(\qone^2\right)^{i(\nu-\nu')-1}e^{i(n-n')\theta}=\delta(\nu-\nu')\delta_{nn'}.
\end{equation}
In this $\ket{n,\nu}$ representation the eigenstates of the LO kernel are 
\begin{equation}
  \label{eq:actionoflok}
  \opk_0\ket{\nu,n} = \asbar\chi_0\left(|n|,\frac{1}{2}+i\nu\right)\ket{\nu,n},
\end{equation}
with $\asbar=\alpha_s N_c/\pi$ and
\begin{equation}
\chi_0(n,\gamma) =2\psi(1)-\psi\left(\gamma+\frac{n}{2}\right)-\psi\left(1-\gamma+\frac{n}{2}\right),
\label{eq:lokernel}
\end{equation}
where $\psi(x)=\Gamma'(x)/\Gamma(x)$, with $\Gamma$ being the Euler gamma function.

Unfortunately, to the best of our knowledge, the azimuthal angle correlation 
between the electron and a forward jet has not been extracted from the HERA 
data so far. For a future comparison with the experimental results in this 
work we implement the same kinematical cuts and constraints as those used 
at HERA. The ZEUS~\cite{Chekanov:2005yb} and H1~\cite{Aktas:2005up} 
collaborations have imposed upper and lower cuts in the transverse 
momentum of the forward jet taking into account the photon virtuality 
$Q^2$. They performed these cuts in order to ensure that both ends of 
the gluon ladder have a similar characteristic transverse scale. 
More in detail, they imposed
\begin{subequations}
\begin{align}
 & {\rm ZEUS:}  & \hspace{-2cm}\frac{1}{2}<\frac{\ktwo^2}{Q^2}<2,\\
 & {\rm H1:}  & \hspace{-2cm}\frac{1}{2}<\frac{\ktwo^2}{Q^2}<5 .
\end{align}
\end{subequations}
These requirements are intended to suppress DGLAP evolution without 
affecting the BFKL dynamics. The implementation in our jet vertex 
of these constraints is straightforward. For the ZEUS condition we 
have
\begin{multline}
  \frac{1}{2}\int d \ktwo^2 \int\ddqtwo \; \braket{n',\nu'}{\qtwo}\bra{\qtwo}{\hat{\Phi}_{\rm jet}}\ket{\ktwo} \\
=: c_2(\nu')\frac{e^{-in'\alpha_2}}{2\pi}= \frac{1}{\sqrt{2}}\frac{1}{\frac{1}{2}+i\nu'}\minusinup{\frac{Q^2}{2}}\left[1-\plusinup{\frac{1}{4}}\right]\frac{e^{-in'\alpha_2}}{2\pi}.
\label{eq:defc2mod}
\end{multline}
In the case of the H1 condition the $1/4$ should be replaced for a $1/10$. For simplicity, in the following, we follow the ZEUS cut.

In Ref.~\cite{Bartels:1996wx} it was shown that the leptonic vertex, 
in our notation, reads
\begin{align}
&\hspace{-1cm} \bra{y,\alpha_1} \oplept\ket{\qone} \non
=& \int dQ^2 \frac{4\alpha^2}{\pi^2N_c\;y\,Q^2}\sum_q e_q^2\int_0^1\int_0^1\frac{ d\xi\, d\zeta}{\xi(1-\xi)Q^2+\zeta(1-\zeta)\qone^2}\non
&\times\Bigg\{\left(\frac{1}{2}-\xi(1-\xi)-\zeta(1-\zeta)+2\xi(1-\xi)\zeta(1-\zeta)\right)y^2\non
&\phantom{\times\Bigg\{}+\Big(1-2\xi(1-\xi)-2\zeta(1-\zeta)+12\xi(1-\xi)\zeta(1-\zeta)\Big)(1-y)\non
&\phantom{\times\Bigg\{}-4\xi(1-\xi)\zeta(1-\zeta)(1-y)\cos\big(2(\theta_1-\alpha_1)\big)\Bigg\}\non
=: & \int dQ^2\left[2a_1^{(0)}(\qone^2,y,Q^2) + 2a_1^{(2)}(\qone^2,y,Q^2)\cos\big(2(\theta_1-\alpha_1)\big)\right],
\end{align}
where $\alpha$ denotes the electromagnetic fine structure constant and $\sum_q e_q^2$ is the sum over the electric charges of the produced quark--antiquark pairs. 

To construct our cross sections we need to find the projection of 
this leptonic impact factor onto the $\ket{\nu,n}$ basis. We obtained
\begin{align}
&  \int\ddqone \bra{y,\alpha_1} \oplept\ket{\qone}\braket{\qone}{\nu,n}\non
%=&\int dQ^2\int\frac{ d\theta_1\,d\qone^2}{2}\left[2a_1^{(0)}(\qone^2,y,Q^2)+2a_1^{(2)}(\qone^2,y,Q^2)\cos\big(2(\theta_1-\alpha_1)\big)\right]\non
%&\hphantom{\int dQ^2\frac{ d\theta_1\,d\qone^2}{2}}\times\frac{1}{\pi\sqrt{2}}\plusinu{\qone^2}e^{in\theta_1}\non
=& \int dQ^2\left[2A_1^{(0)}\left(\nu,y,Q^2\right)+A_1^{(2)}\left(\nu,y,Q^2\right)\left(\delta_{n,-2}e^{-2i\alpha_1}+\delta_{n,2}e^{2i\alpha_1}\right)\right], \label{eq:leptonicif}
\end{align}
with
\begin{equation}
  A_1^{(n)}\left(\nu,y,Q^2\right) = \frac{1}{\sqrt{2}}\int d\qone^2\; a_1^{(n)}(\qone^2,y,Q^2)\plusinu{\qone^2} .
\end{equation}
To calculate these coefficients we need integrals of the type
\begin{multline}
  \int_0^1 d\xi\,(\xi(1-\xi))^{t_\xi}\int_0^1d\zeta\,(\zeta(1-\zeta))^{t_\zeta}\int_0^\infty d\qone^2\frac{\plusinu{\qone^2}}{\xi(1-\xi)Q^2+\zeta(1-\zeta)\qone^2}=\\
\frac{\pi \plusinu{Q^2}}{\cosh (\pi\nu)} B\left(\frac{1}{2}+t_\xi+i\nu,\frac{1}{2}+t_\xi+i\nu\right) B\left(\frac{1}{2}+t_\zeta-i\nu,\frac{1}{2}+t_\zeta-i\nu\right),\label{eq:leptimpproj}
\end{multline}
with $B$ representing the Euler beta function. Using this formula for
 $t_{\xi/\zeta}\in\{0,1\}$ we obtained 
\begin{subequations}
\begin{align}
  A_1^{(0)}\left(\nu,y,Q^2\right) =& \phantom{\times}\frac{\alpha^2\sqrt{2}}{y\,N_c}\plusinu{Q^2}\sum_q e_q^2\frac{1}{16\nu(\nu^2+1)}\frac{\tanh(\pi\nu)}{\cosh(\pi\nu)}\non
&\times\left(\frac{4\nu^2+9}{2}y^2+(12\nu^2+11)(1-y)\right)\label{eq:defA0},\\ 
  A_1^{(2)}\left(\nu,y,Q^2\right) =& \phantom{\times}\frac{\alpha^2\sqrt{2}}{y\,N_c}\plusinu{Q^2}\sum_q e_q^2\frac{1}{16\nu(\nu^2+1)}\frac{\tanh(\pi\nu)}{\cosh(\pi\nu)}\non
&\times\left(-(4\nu^2+1)(1-y)\right)\label{eq:defA2} .
\end{align}
\end{subequations}

The last piece needed to complete Eq.~\eqref{eq:oppartonicep2} is the 
gluon Green's function, which can be written as 
\begin{equation}
\bra{n,\nu}\opf\ket{\nu',n'} =  \int\frac{d\omega}{2\pi i}\,\bra{n,\nu}\opf_\omega\ket{\nu',n'} e^{\omega Y} 
= e^{\chi\left(|n|,\frac{1}{2}+i\nu,\asbar\right) Y}\;\delta(\nu-\nu')\delta_{nn'}  ,
\label{eq:fnnu}
\end{equation}
with the eigenvalue of the BFKL kernel being
\begin{multline}
  \chi\Big(n,\frac{1}{2}+i\nu,\asbar\Big) = \asbar \chi_0\Big(n,\frac{1}{2}+i\nu\Big)\\
+\asbar^2\left(\chi_1\Big(n,\frac{1}{2}+i\nu\Big)-\frac{\beta_0}{8N_c}
%\left[i\frac{\partial}{\partial\nu}\ln\frac{A^{(n)}\left(\nu,y,Q^2\right)}{c_2(\nu)}\right]
\chi_0\Big(n,\frac{1}{2}+i\nu\Big)h_{\rm rc}^{(n)}\left(\nu,y,Q^2\right)\right)
.\label{eq:chi}
\end{multline}
The action of the scale invariant sector of the NLO correction, given by the function $\chi_1(n,\gamma)$, was calculated in Ref.~\cite{Kotikov:2000pm}. 
The last term in this equation stems from the scale dependent part of the NLO kernel, i.e. from the running of the coupling. Its explicit form, in our representation,  depends on the impact factors as given below and is discussed in more detail in Refs.~\cite{Vera:2006un,Schwennsen:2007hs}.
\begin{subequations}
\label{eq:impcontributiondis}
\begin{align}
%  i\frac{\partial}{\partial\nu}\ln\frac{A_1^{(0)}\left(\nu,y,Q^2\right)}{c_2(\nu)} 
h_{\rm rc}^{(0)}\left(\nu,y,Q^2\right)=& %-2\ln Q^2
-\frac{1}{2\nu^2+\frac{1}{2}}+\frac{3\ln (2)}{5-4\cos(\nu\,\ln 4)}\non
&-i\Bigg[ \pi\frac{\cosh(2\pi\nu)-3}{\sinh(2\pi\nu)}+\frac{4\sin(\nu\,\ln 4)\ln (2)}{5-4\cos(\nu\,\ln 4)} \non
&\hphantom{-i\Bigg[}-8\nu\frac{y^2+6(1-y)}{9y^2+22(1-y)+4(y^2+6(1-y))\nu^2} \non
&\hphantom{-i\Bigg[}  +\frac{3\nu^2+1}{\nu(\nu^2+1)}-\frac{\nu}{\nu^2+\frac{1}{4}} \Bigg] , \\
%  i\frac{\partial}{\partial\nu}\ln\frac{A_1^{(2)}\left(\nu,y,Q^2\right)}{c_2(\nu)} 
h_{\rm rc}^{(2)}\left(\nu,y,Q^2\right)=& %-2\ln Q^2
-\frac{1}{2\nu^2+\frac{1}{2}}+\frac{3\ln (2)}{5-4\cos(\nu\,\ln 4)}\non
&-i\Bigg[\pi\frac{\cosh(2\pi\nu)-3}{\sinh(2\pi\nu)}+\frac{4\sin(\nu\,\ln 4)\ln (2)}{5-4\cos(\nu\,\ln 4)}\non
&\hphantom{-i\Bigg[}+ \frac{3\nu^2+1}{\nu(\nu^2+1)}-\frac{3\nu}{\nu^2+\frac{1}{4}} \Bigg] .
\end{align}
\end{subequations}

Blending together the leptonic vertex of Eq.~\eqref{eq:leptonicif}, the 
Green's function of Eq.~\eqref{eq:fnnu} and the forward jet vertex 
given in Eq.~\eqref{eq:defc2mod}, we obtain the cross 
section of Eq.~\eqref{eq:oppartonicep2}, which can be expressed in 
differential form as 
\begin{multline}
  \frac{d\hat\sigma}{dy\,dQ^2\,d\phi} = 
\frac{\pi^2\asbar^2}{2}\int d\nu \int d\nu' \Bigg[A_1^{(0)}\left(\nu,y,Q^2\right)\bra{0,\nu}\opf\ket{\nu',0}c_2(\nu')\\
+A_1^{(2)}\left(\nu,y,Q^2\right)\bra{2,\nu}\opf\ket{\nu',2}c_2(\nu')\cos\,2\phi\Bigg],\label{eq:oppartonicep3}
\end{multline}
where we have introduced the azimuthal angle $\phi=\alpha_2-\alpha_1$ between the electron and the forward jet. We have also made use of the relation $\bra{n,\nu}\opf\ket{\nu',n}=\bra{-n,\nu}\opf\ket{\nu',-n}$.

It is more convenient to write Eq.~\eqref{eq:oppartonicep3} as 
\begin{equation}
  \frac{d\hat{\sigma}}{d\phi\;dy\;dQ^2} = \frac{\pi^2\asbar^2}{2}\left[B^{(0)}\left(y,Q^2,Y\right) +B^{(2)}\left(y,Q^2,Y\right) \cos 2\phi\right],
\end{equation}
where the coefficients $B^{(n)}$ at LO read
\begin{equation}
  B^{(n)}_{\rm LO}\left(y,Q^2,Y\right)= \int d\nu \;A^{(n)}\left(\nu,y,Q^2\right)\,c_2(\nu) e^{Y\asbar \chi_0\left(\left|n\right|,\nu\right)},
\end{equation}
and at NLO:
\begin{multline}
 B^{(n)}_{\rm NLO}\left(y,Q^2,Y\right)= \int d\nu \;A^{(n)}\left(\nu,y,Q^2\right)\,c_2(\nu)\\
\times e^{{\bar \alpha}_s (Q^2){\rm Y} \left(\chi_0\left(\left|n\right|,\nu\right)+{\bar \alpha}_s  (Q^2) \left(\chi_1\left(\left|n\right|,\nu\right)-\frac{\beta_0}{8 N_c} \chi_0(n,\frac{1}{2}+i\nu)h_{\rm rc}^{(n)}(\nu,y,Q^2)\right)\right)}.
\end{multline}

The BFKL resummation presents an instability in transverse momentum 
space when the NLO corrections are taken into account~\cite{Ross:1998xw}.
A prescription to increase the convergence of the perturbative expansion 
is to improve the original calculation by imposing compatibility of the 
scattering amplitudes with the collinear limit dominated by renormalization group evolution~\cite{Salam:1998tj,Vera:2005jt}. In recent 
publications~\cite{Vera:2007kn,Schwennsen:2007hs} we have introduced  
these collinear improvements to describe azimuthal angle dependences in 
the context of Mueller--Navelet jets. Our results were later on 
reproduced in Ref.~\cite{Royon2}. 

From a technical point of view,  
the collinearly--improved kernel of Ref.~\cite{Vera:2007kn} differs 
from the one needed in the DIS case only in the term due to 
the running of the coupling in Eq.~\eqref{eq:chi}. This contribution 
changes the single and double poles of the original kernel in the form
\begin{align}
a_0\quad\rightarrow\quad &a_0 -\frac{\beta_0}{8N_c}\left(\frac{7}{6}+\frac{1-y}{y\left(\frac{y}{2}-1\right)+1}\right), \\
a_2\quad\rightarrow\quad &a_2 -\frac{\beta_0}{8N_c}\left(\frac{107}{30}+\frac{5\ln 2}{3}\right), \\
b_n\quad\rightarrow\quad &b_n +\frac{\beta_0}{4N_c} .
\end{align}
These equations hence replace Eqs.~(24, 25) of Ref.~\cite{Vera:2007kn}.

%For the coefficients with resummed kernel we can write accordingly
%\begin{equation}
%  B^{(n)}_{\rm resum}\left(y,Q^2,Y\right)= \int d\nu \;A^{(n)}\left(\nu,y,Q^2\right)\,c_2(\nu) e^{Y\,\omega^{\rm resum}\left(\left|n\right|,\nu\right)}.
%\end{equation}
%and $A_1{(0/2)}$ are given in Eqs.~(\ref{eq:defA0}, \ref{eq:defA0}) and for $c_2(\nu)$ we use the modified one of Eq.~\eqref{eq:defc2mod}.

\section{Phenomenology}

Besides the particular experimental cuts in the forward jet taken into 
account when calculating the jet vertex of Eq.~\eqref{eq:defc2mod}, we 
also used the following experimentally motivated~\cite{Didar:private} 
constraints in the leptonic sector:
\begin{eqnarray}
&&20 ~{\rm GeV}^2  < Q^2   < 100 ~{\rm GeV}^2, \nonumber\\
&&0.05 < y < 0.7 , \nonumber\\
&&5\cdot 10^{-3} > x_{\rm Bj} > 4\cdot 10^{-4}. 
\label{eq:heracuts}
\end{eqnarray}
The final expression for the cross section at hadronic level reads
\begin{equation}
  \label{eq:disfinal}
  \frac{d\sigma}{dY\;d\phi} =: C_0(Y)+C_2(Y)\cos 2\phi ,
\end{equation}
with
\begin{equation}
  C_n(Y) = \frac{\pi^2\asbar^2}{2}\int_{\rm cuts} \hspace{-.3cm}dx_{\rm FJ}\,dQ^2\,dy\,f_{\rm eff}(x_{\rm FJ},Q^2) B^{(n)}(y,Q^2,Y) \delta\left(x_{\rm FJ}-\frac{Q^2 e^Y}{ys}\right),
\end{equation}
where we performed the convolution with the effective parton 
distribution of  Eq.~\eqref{eq:feff}. The index in the integral sign 
refers to the particular cuts of Eq.~\eqref{eq:heracuts}. The 
integration over the longitudinal momentum fraction $x_{\rm FJ}$ of the 
forward jet involves a delta function fixing the rapidity 
$Y=\ln x_{\rm FJ}/x_{\rm Bj}$. It is noteworthy that any additional 
experimental upper cut on $x_{\rm FJ}$ would modify the 
coefficients $C_n$, with a negligible change in their ratios.

\begin{figure}[htbp]
  \centering
  \includegraphics[width=11cm]{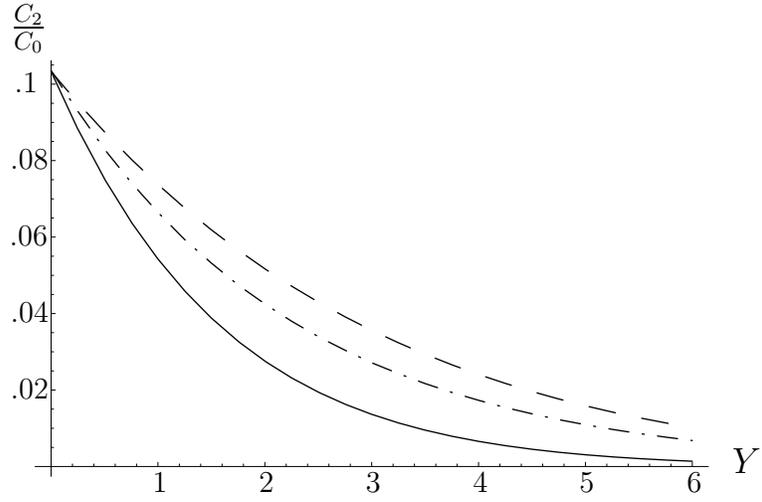}
  \caption{$<\cos 2\phi>$ at the $ep$ collider HERA at leading (solid), next to leading order (dashed), and for resummed kernel (dash-dotted). }
  \label{fig:hera1}
\end{figure}

Since the structure of the electron vertex singles out the components 
with conformal spin 0 and 2, the number of observables related to the 
azimuthal angle dependence is limited when compared to the 
Mueller--Navelet case. The most relevant observable is the dependence 
of the average $<\cos 2\phi> = C_2/C_0$ with the rapidity difference 
between the forward jet and outgoing lepton. It is natural to expect 
that the forward jet will be more decorrelated from the leptonic system 
as the rapidity difference is larger since the phase space for further 
gluon emission opens up. This is indeed what we observe in our 
numerical results shown in Fig.~\ref{fig:hera1}. We find similar results 
to the Mueller--Navelet jets case where the most reliable calculation 
is that with a collinearly--improved kernel. The main effect of the 
higher order corrections is to increase the azimuthal angle correlation 
for a given rapidity difference, while keeping the decrease of the 
correlation as $Y$ grows.
It is interesting to point out that, even for very small $Y$, the 
inclusive quark--antiquark pair (produced to couple the electron 
to the gluon evolution) generates in the case of no gluon emission some 
angular decorrelation between the forward jet and the electron. 

\begin{figure}[htbp]
  \centering
  \includegraphics[width=11cm]{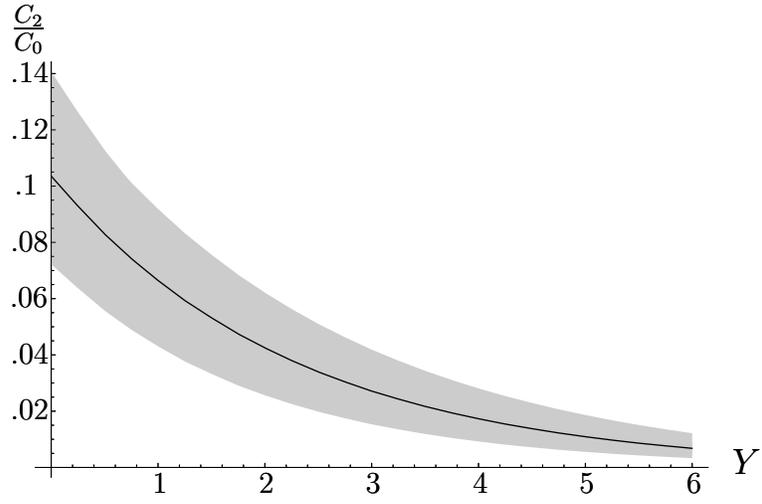}
  \caption{$<\cos 2\phi>$ at the $ep$ collider HERA for resummed kernel (dash-dotted). The gray band reflects the uncertainty in $s_0$ and in the renormalization scale $\mu$.}
  \label{fig:heras0}
\end{figure}

Finally, we estimate the theoretical uncertainties derived from not 
including the NLO impact factors by varying the scale $s_0$, and those 
related to the running of the coupling by doing the same with the 
renormalization scale $\mu$. The range of variation in both parameters  
is between $1/2$ and 2 and the result is shown in Fig.~\ref{fig:heras0}.
%In Fig.~\ref{fig:hera2} it can be seen that  the theoretical uncertainty is larger for smaller rapidity difference,  and rapidly reduces as we move to larger $Y$.

\begin{figure}[htbp]
  \centering
  \includegraphics[width=11cm]{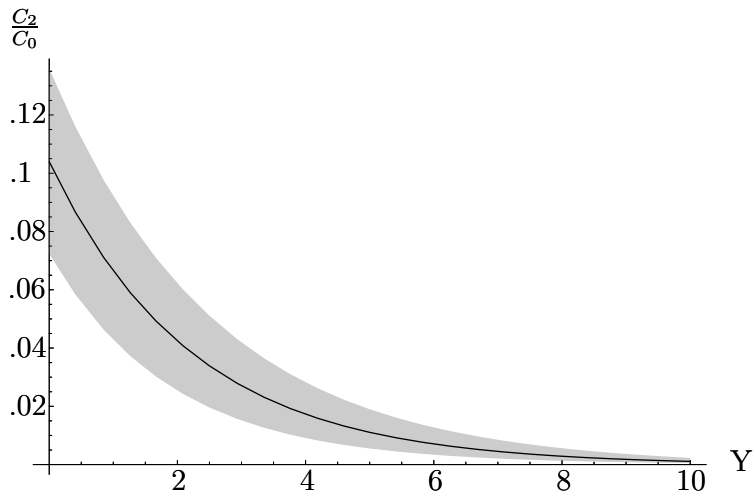}
  \caption{$<\cos 2\phi>$ at a possible  $ep$ collider LHeC with a gray band
reflecting the uncertainty in $s_0$ and in the renormalization scale
$\mu$.}
\label{fig:hera2}
\end{figure}

At present, the data taken at the HERA collider provide the only possibility to experimentally test our prediction. Nevertheless there are proposals to upgrade the Large Hadron Collider (LHC) at CERN to a Large Electron Hadron Collider \cite{Dainton:2006wd}. The projected center of mass energy $\sqrt{s}=1.4~{\rm TeV}$ is more than four times bigger than at HERA and would allow for a larger rapidity separation between the electron and the forward jet. We use the same cuts as for HERA (Eq.~\eqref{eq:heracuts}) apart from an adjusted lower bound for $x_{\rm Bj}$ of $2\cdot 10^{-5}$. Fig.~\ref{fig:hera2} is a plot of our results, 
which are very similar to those presented in Fig.~\ref{fig:heras0}.

\section{Conclusions}

We have studied the effect of higher order corrections to the BFKL 
equation on the angular decorrelation of forward jets in Deep Inelastic 
Scattering. The effect of these additional terms is similar to the 
previously studied case of Mueller--Navelet jets at hadron colliders. 
As the rapidity difference between the outgoing lepton and the forward 
jet increases, the two systems decorrelate in azimuthal angle due to 
the extra emission of soft gluons and higher order terms largely increase the 
amount of correlation when compared to the leading order calculations. 
It would be very interesting to 
extract this dependence from the HERA data with forward jets and 
study how important BFKL effects are for this observable. If the 
experience at the Tevatron is valid in this case, the BFKL prediction 
will probably lie below the data, with a gradual improvement of the fits 
as the rapidity difference increases. 

\begin{flushleft}
{\bf \large Acknowledgments}
\end{flushleft}
We would like to thank J.~Bartels, D.~Dobur, H.~Jung, L.~Motyka and 
J.~Terr{\' o}n for very interesting discussions.  F.S. is supported by the 
Graduiertenkolleg ``Zuk\"unftige Entwicklungen in der Teilchenphysik''.

%\bibliographystyle{h-physrev3}
%\bibliography{literaturgesamt}

\begin{thebibliography}{10}

\bibitem{Vera:2006un}
  A.~Sabio~Vera, Nucl.\ Phys.\  B {\bf 746} (2006) 1. 
  %%CITATION = NUPHA,B746,1;%%

\bibitem{Vera:2007kn}
  A.~Sabio~Vera and F.~Schwennsen, Nucl.\ Phys.\  B {\bf 776} (2007) 170.
  %%CITATION = HEP-PH/0702158;%%

\bibitem{Schwennsen:2007hs}
F.~Schwennsen,
\newblock DESY-THESIS-2007-001, hep-ph/0703198.

\bibitem{BFKL}
V.S.~Fadin, E.A.~Kuraev and L.N.~Lipatov, Phys. Lett. {\bf B60} (1975) 50;
E.A.~Kuraev, L.N.~Lipatov and V.S.~Fadin, Zh. Eksp. Teor. Fiz. {\bf 71} (1976)
840 [Sov. Phys. JETP {\bf 44} (1976) 443]; {\bf 72} (1977) 377 
[{\bf 45} (1977) 199];
Ya.Ya.~Balitskii and L.N.~Lipatov, Sov. J. Nucl. Phys. {\bf 28} (1978) 822.

\bibitem{NLA-kernel}
V.S.~Fadin and L.N.~Lipatov, Phys. Lett. {\bf B429} (1998) 127;
M.~Ciafaloni and G.~Camici, Phys. Lett. {\bf B430} (1998) 349.

\bibitem{Bartels:1996wx}
J.~Bartels, V.~Del~Duca, and M.~Wusthoff,
\newblock Z. Phys. {\bf C76}, 75 (1997).
%%CITATION = HEP-PH 9610450;%%

\bibitem{Kepka}
  O.~Kepka, C.~Royon, C.~Marquet and R.~Peschanski, hep-ph/0609299, hep-ph/0612261.

\bibitem{ImpactFactorsAll}
J.~Bartels, S.~Gieseke and C.~F. Qiao, Phys. Rev. {\bf D63}, 056014 (2001);
%%CITATION = HEP-PH 0009102;%%
J.~Bartels, S.~Gieseke and A.~Kyrieleis, Phys. Rev. {\bf D65}, 014006 (2002);
%%CITATION = HEP-PH 0107152;%%
J.~Bartels, D.~Colferai, S.~Gieseke and A.~Kyrieleis, Phys. Rev. {\bf D66}, 094017 (2002); 
%%CITATION = HEP-PH 0208130;%%
J.~Bartels and A.~Kyrieleis, Phys. Rev. {\bf D70}, 114003 (2004);
%%CITATION = HEP-PH 0407051;%%
G.~Chachamis, DESY-THESIS-2006-031;
%%CITATION = DESY-THESIS-2006-031;%%
V.~S. Fadin, D.~Y. Ivanov and M.~I. Kotsky, Phys. Atom. Nucl. {\bf 65}, 1513 (2002);
%%CITATION = HEP-PH 0106099;%%
V.~S. Fadin, D.~Y. Ivanov and M.~I. Kotsky, Nucl. Phys. {\bf B658}, 156 (2003).
%%CITATION = HEP-PH 0210406;%%


\bibitem{NLOjetvertex}
J.~Bartels, D.~Colferai and G.~P.~Vacca, Eur.\ Phys.\ J.\  C {\bf 24} (2002) 83;  %%CITATION = EPHJA,C24,83;%%
Eur.\ Phys.\ J.\  C {\bf 29} (2003) 235.
  %%CITATION = EPHJA,C29,235;%%


\bibitem{Combridge:1983jn}
B.~L. Combridge and C.~J. Maxwell, Nucl. Phys. {\bf B239}, 429 (1984).
%%CITATION = NUPHA,B239,429;%%

\bibitem{Chekanov:2005yb}
ZEUS, S.~Chekanov {\em et~al.}, Phys. Lett. {\bf B632}, 13 (2006).
%%CITATION = HEP-EX 0502029;%%

\bibitem{Aktas:2005up}
H1, A.~Aktas {\em et~al.}, Eur. Phys. J. {\bf C46}, 27 (2006).
%%CITATION = HEP-EX 0508055;%%

\bibitem{Kotikov:2000pm}
A.~V. Kotikov and L.~N. Lipatov, Nucl. Phys. {\bf B582}, 19 (2000).
%%CITATION = HEP-PH 0004008;%%

\bibitem{Ross:1998xw}
D.~A. Ross, Phys. Lett. {\bf B431}, 161 (1998).
%%CITATION = HEP-PH 9804332;%%

\bibitem{Salam:1998tj}
G.~P. Salam, JHEP {\bf 07}, 019 (1998).
%%CITATION = HEP-PH 9806482;%%

\bibitem{Vera:2005jt}
A.~Sabio~Vera, Nucl. Phys. {\bf B722}, 65 (2005).
%%CITATION = HEP-PH 0505128;%%

\bibitem{Royon2}
  C.~Marquet and C.~Royon, arXiv:0704.3409.

\bibitem{Didar:private}
D.~Dobur,
\newblock  {private communications}.

\bibitem{Dainton:2006wd}
  J.~B.~Dainton, M.~Klein, P.~Newman, E.~Perez and F.~Willeke,
  %``Deep inelastic electron nucleon scattering at the LHC,''
  JINST {\bf 1} (2006) P10001.
  %%CITATION = JINST,1,P10001;%%










\end{thebibliography}

\end{document}